\documentclass[12pt]{iopart}

\usepackage[utf8]{inputenc}
\usepackage[T1]{fontenc}
\usepackage{amstext}
\usepackage{array}
\usepackage{multirow}
\usepackage{multicol}
\usepackage{booktabs}
\usepackage{graphicx}
\graphicspath{ {./images/} }
\usepackage[english]{babel}
\usepackage{amsmath}
\usepackage{bm}
\usepackage{algorithm2e}
\usepackage{lineno}
\usepackage{physics}
\usepackage{empheq}
\usepackage{cite}
\usepackage{eufrak}
\usepackage{lipsum}
\usepackage{float}
\usepackage[colorlinks, citecolor = blue, urlcolor = blue]{hyperref}
\hypersetup{breaklinks=true}
\usepackage{url}
\usepackage{booktabs}

\begin{document}
\global\long\def\a{\alpha}%

\global\long\def\s{\sigma}%

\global\long\def\p{\psi}%

\global\long\def\l{\lambda}%

\global\long\def\o{\phi_{k}}%

\global\long\def\r{\rho_{k}}%

\global\long\def\e{\xi}%

\title[]{Quantum theory of orbital angular momentum  in spatiotemporal optical vortices}

\author{Pronoy Das, Sathwik Bharadwaj and Zubin Jacob$^*$}

\address{Elmore Family School of Electrical and Computer Engineering, Birck Nanotechnology Center, Purdue University, West Lafayette, IN 47907, United States of America}
\ead{zjacob@purdue.edu$^*$}

\vspace{10pt}
\begin{indented}
\item[]March 2024
\end{indented}

\begin{abstract}
Spatiotemporal Optical Vortices (STOVs) are structured electromagnetic fields propagating in free space with phase singularities in the space-time domain. Depending on the tilt of the helical phase front, STOVs can carry both longitudinal and transverse orbital angular momentum (OAM). Although STOVs have gained significant interest in the recent years, the current understanding is limited to the semi-classical picture. Here, we develop a quantum theory for STOVs with an arbitrary tilt, extending beyond the paraxial limit. We demonstrate that quantum STOV states, such as Fock and coherent twisted photon pulses, display non-vanishing longitudinal OAM fluctuations that are absent in conventional monochromatic twisted pulses. We show that these quantum fluctuations exhibit a unique texture, i.e. a spatial distribution which can be used to experimentally isolate these quantum effects. Our findings represent a step towards the exploitation of quantum effects of structured light for various applications such as OAM-based encoding protocols and platforms to explore novel light-matter interaction in 2D material systems.
\end{abstract}

%
%
%
%
%

\section{Introduction}

\begin{figure}[!ht]
    \centering
    \includegraphics[width=1\textwidth]{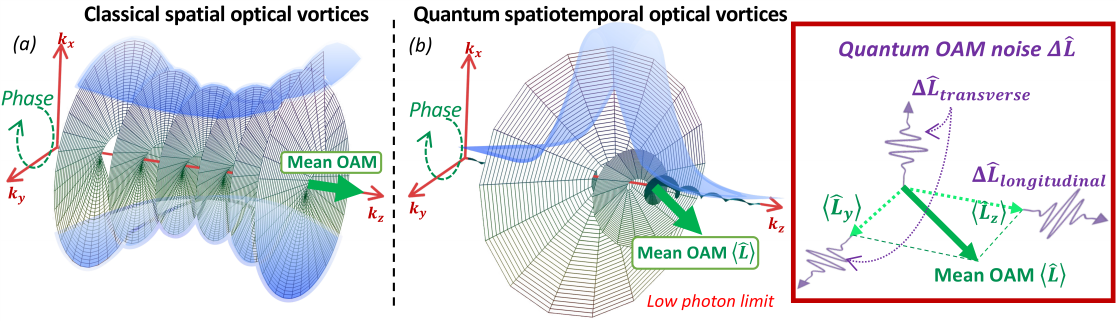}
    \caption{\textbf{Difference between a semi-classical twisted beam and a quantum spatiotemporal twisted pulse:} (a) a conventional monochromatic twisted beam with high photon count, which captures the mean OAM along $k_z$ of the beam (b) a quantum polychromatic spatiotemporal pulse with low photon count, with both transverse and longitudinal OAM components. The quantum effects of photon statistics leads to fluctuations in both the transverse and longitudinal directions of the OAM ($\Delta {\hat{L}}_i = \sqrt{\langle {\hat{L}}_i^2\rangle - \langle {\hat{L}}_i\rangle^2}$).}
    \label{F1}
\end{figure}

Conventional monochromatic twisted light pulses or optical vortices have a vanishing phase intensity at the center, with a helical phase-front winding around the propagation axis (Fig. \ref{F1}(a)) \cite{Shen2019,  Dunlop2017, Zhuang2004, Padgett2011, Macdonald2002, Miao2016, Mair2001, Barreiro2008}. Such optical vortices have a time-independent phase singularity at the center and carry an orbital angular momentum (OAM) along the longitudinal direction (i.e., the direction of propagation). Recently, there have been rapid advancements in a new frontier of twisted pulses, known as spatiotemporal optical vortices (STOVs), that can carry the OAM in any arbitrary direction (Fig. \ref{F1}(b)) \cite{DROR2011526, Bliokh2012, Jhajj2016, Zhao2020, Zhang_2022, Hancock2021, Forbes2021, Wan2023}. These vortices are a generalized form of monochromatic twisted pulses in the space-time domain and feature time-dependent phase singularities. Unlocking such new degrees of freedom in the OAM results in various novel applications of STOVs, such as encoding data states by multiplexing in communication \cite{Ni2021} and trapping, manipulating, and even transporting nanoparticles using phase singularities in the helical phase \cite{Stilgoe2022}. However, the current frontier in the field of STOVs is limited by the semi-classical understanding of their behavior, which fails to capture their quantum nature at the few photon level.

Recently, quantum fluctuations in the OAM of a conventional monochromatic quantum twisted pulse have been studied theoretically \cite{Yang2021}. However, they are limited to the transverse plane, thus failing to provide a three-dimensional (3D) quantum picture for such fluctuations. Extending beyond the limitations of the conventional twisted pulse and unlocking additional OAM degrees of freedom can describe novel subatomic phenomena in 2D material systems \cite{Crimin2019}. Moreover, using an arbitrary OAM photon from a deterministic single-photon source (e.g., a Fock-state pulse from semiconductor quantum dots), we can generate higher-dimensional quantum states using OAM-based encoding protocols \cite{Suprano2023,Wang2012}. This provides new avenues for the scalability of qudit-based systems and improves the security of communication protocols as it expands the amount of information supported by a single STOV photon.

In this paper, we develop a quantum theory of the orbital angular momentum of spatiotemporal twisted pulses with arbitrary tilt. We consider the particular cases of Fock and coherent state photon pulses, however the proposed theory is valid beyond the aforementioned states. Moreover, structured light pulses in the low photon limit experience fundamental quantum fluctuations, such as intensity or shot noise and optical phase noise \cite{Henry1996,Clerk2010}. Unlike a monochromatic optical vortex, we show that quantum spatiotemporal vortices inherit a non-vanishing OAM noise in the longitudinal direction. We show that these fluctuations have a unique spatial distribution, which evolves with time. We define this texture of local OAM fluctuations as the OAM noise density. We note that our work is universally applicable to any arbitrary quantum optical vortex pulse and can be extended beyond the paraxial limit.

\section{Results}

\subsection{Wavefunction of quantum spatiotemporal twisted pulse}

In this section, we define the wavefunction of the quantum STOV. The single-photon wave-packet creation operator $\hat{a}^\dagger_{\xi,\lambda}$ for the twisted pulse can be written as a coherent superposition of the plane-wave modes $\hat{a}^\dagger_{k,\lambda}$, given by
\begin{equation}
    \hat{a}^\dagger_{\xi,\lambda}=\int d^3k \xi_l(\bm{k}) \hat{a}^\dagger_{k,\lambda} , \nonumber
\end{equation}

\noindent where $\xi_l(\bm{k})$ is the spectral amplitude function (SAF) of the twisted pulse. This amplitude function determines both the structural and quantum attributes of the quantum pulse.

\begin{figure}[ht]
    \centering
    \includegraphics[width=0.8\textwidth]{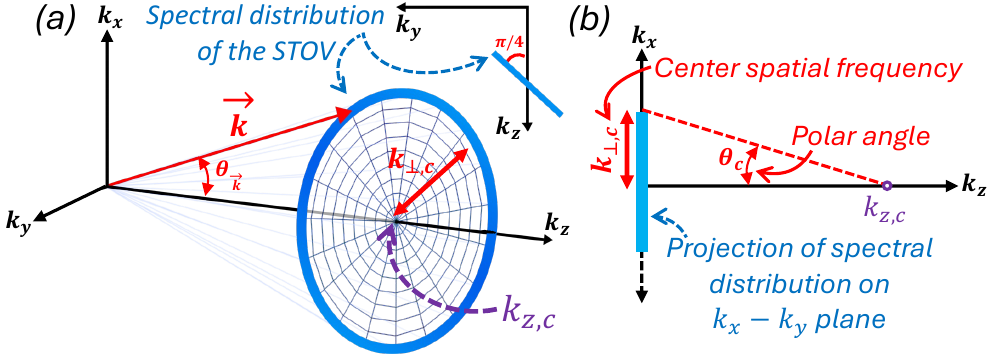}
    \caption{\textbf{Definition of the polar angle $\theta_c$ and center spatial frequency $k_{\perp,c}$}: (a) Schematic of the spectral distribution of the spatiotemporal Bessel-Gaussian pulse. (b) Projection of the spectral distribution on the $k_x-k_y$ plane, viewed on the $k_y-k_z$ plane. $\theta_c$ is the polar angle for this twisted pulse.}
    \label{AF1}
\end{figure}

In Appendix B, we explicitly derive the spectral amplitude function (SAF) of the spatiotemporal twisted pulse, given by
\begin{eqnarray}
    \label{eq1}
    &\e_{l}(\r,k_{z},\o)\\ \nonumber
    &=  \frac{1}{\sqrt{2\pi}}\Big(\frac{2\s_{z}^{2}}{\pi}\Big)^{\frac{1}{4}}\Big(\frac{2\sigma_{\rho}^{2}}{\pi k_{\perp,c}^{2}}\Big)^{\frac{1}{4}}\\ \nonumber
    & \times\exp\Bigg[-\s_{z}^{2}(-\r\sin\o\sin\theta+(k_{z}-k_{z,c})\cos\theta)^{2}\Bigg]\\ \nonumber
     &\times\exp\Bigg[-\s_{\rho}^{2}(\sqrt{(\r\cos\o)^{2}+(\r\sin\o\cos\theta+(k_{z}-k_{z,c})\sin\theta)^{2}}-k_{\perp,c})^{2}\\ \nonumber
    & \qquad+il\tan^{-1}\frac{\r\sin\o\cos\theta+(k_{z}-k_{z,c})\sin\theta}{\r\cos\o}\Bigg],
\end{eqnarray}

\noindent where $\r,k_{z},\o$ represent the radial distance, axial coordinate and azimuth for the cylindrical coordinates in \textit{k}-space. $\sigma_z$ and $\sigma_\rho$ are the width of the Gaussian functions which characterizes the envelope of the pulse, with $k_{z,c}$ and $k_{\perp,c}$ as their respective center spatial frequencies. The phase-front of the STOV has a twist $\theta$ along a chosen axis in the $k_x-k_y$ plane. Also, $l$ represents the helical phase index of the SAF. In Fig. \ref{AF1}, we have shown a schematic of the spectral distribution of the spatiotemporal twisted Bessel-Gaussian pulse.

The first Gaussian function with a width $1/\sigma_z$ characterizes the envelope of the wavefront. The pulse length $\tau$ can be explicitly determined by the relation $\sigma_z=c\tau$. Although the SAF of a Bessel pulse in $k-$space contains the delta function $\delta(\theta_k-\theta_c)$ \cite{Arnaut2000,Jentschura2011} ($\theta_c$ is the polar angle, Fig. \ref{AF1}) to characterize the SAF distribution in the $k_x-k_y$ plane, this poses a problem with the normalization of the wavefunction of this quantum pulse \cite{Yang2021}. In order to circumnavigate this issue, we replace this delta function with a second Gaussian of width $1/\sigma_\rho$. The center spatial frequency of this Gaussian is linked to $k_{z,c}$ through the expression $k_{\perp,c}=k_{z,c}\tan\theta_c$. 

The imaginary component in the exponential function provides the phase-front of the twisted beam its signature helical nature. The tilt angle $\theta$ of this helical structure plays a decisive role in determining the vector nature of the mean OAM \cite{Allen1992}. Without loss of generality, we have chosen $\theta$ to be equal to $-\pi/4$. We note that that setting $\theta=(2m+1)\pi/2$ results in complete longitudinal OAM, similar to a conventional monochromatic twisted quantum Bessel-Gaussian pulse \cite{Yang2021}. Conversely, we can get a purely transverse OAM with $\theta=m\pi$ \cite{Chong2020} ($m$ is an integer). 

We introduce the definition of transverse center spatial frequency $k_{\perp,c}$, and consequently polar angle for the spatiotemporal pulse (Fig. \ref{AF1}), since the projection of the STOV's spectral distribution onto the $k_x-k_y$ plane is an ellipse rather than a circle. However, our definition of the polar angle is geometrically consistent with the widely-known definition for a spatial beam. Thus, the stated relation $k_{\perp,c}=\tan(\theta_c)k_{z,c}$ remains valid, and we can derive $k_{\perp,z}$ from the major axis of this projected ellipse on the $k_x-k_y$ plane.

The SAF $\xi_l(\bm{k})$ must satisfy the normalization condition to ensure that the single-photon creation (annihilation) operators $\hat{a}^\dagger_{\xi,\lambda}(\hat{a}_{\xi,\lambda})$ satisfy the bosonic commutation relations
\begin{equation}
    [\hat{a}_{\xi,\lambda},\hat{a}^\dagger_{\xi,\lambda'}] = \delta_{\lambda,\lambda'}\qquad \text{and} \qquad[\hat{a}_{\xi,\lambda},\hat{a}_{\xi,\lambda'}]=[\hat{a}^\dagger_{\xi,\lambda},\hat{a}^\dagger_{\xi,\lambda'}]=0. \nonumber
\end{equation}

\noindent We numerically computed the quantity $\int d^3k |\xi_l(\bm{k})|^2$, which yields a value of 1 (error $=4\times10^{-6}$), thus proving that our SAF is normalizable.

\subsection{Photonic Fock state and coherent state}

We can construct the wave function for any arbitrary quantum pulse using the bosonic ladder operators defined in the previous subsection. Although the theory is applicable for any quantum photonic state, we demonstrate our proposed framework for the two well-known states: the photonic Fock and the coherent states, given by
\begin{eqnarray*}
\centering
    \text{Fock-state:}&\;|n_{\xi,\lambda} \rangle= \frac{1}{\sqrt{n!}}\Big(\hat{a}_{\xi,\lambda}^\dagger \Big)^n |0\rangle\\
    \text{Coherent state:}&\;|\alpha_{\xi,\lambda} \rangle = e^{-\Tilde{n}/2}\sum_{k=0}^\infty \frac{\alpha^k}{\sqrt{k!}}|n_{\xi,\lambda}\rangle
\end{eqnarray*}

\noindent where $n$ is the number of photons in our Fock state pulse, and $\Tilde{n}=|\alpha|^2$ represents the average photon number of the coherent state pulse.


\subsection{Quantum OAM operator and commutation relations}

In this section, we define the quantum operator for the OAM and the corresponding commutation relations. We can derive the expression for the quantum OAM operator for the photon

\begin{equation}
    \hat{\bm{L}} =-i\hbar \int d^3k \sum_{\lambda=\pm 1}[\hat{a}_{k,\lambda}^\dagger(\boldsymbol{k \times \nabla_k})\hat{a}_{k,\lambda}],
    \label{eq2}
\end{equation}

\noindent by applying a Lorentz gauge-fixing condition on the gauge-dependent quantum OAM operator $\hat{\Tilde{\bm{L}}} =-\int d^{3}x[\hat{\pi}^{j}(\boldsymbol{x\times\nabla})\hat{A}^j]$ \cite{Yang2022}. Here, $\hat{A}$ and $\hat{\pi}$ represent the vector potential and the conjugate momentum operators respectively.

These commutator relations for the quantum OAM operator for the (spin-1) photon (or Maxwell fields)
\begin{equation}
    [\hat{L}_i,\hat{L}_j]=i\hbar\epsilon_{ijk}\hat{L}_k,
    \label{eq3}
\end{equation}

\noindent mirror the well-known OAM relations for the (spin-$1/2$) Dirac fields \cite{Yang2022}, resolving the dilemma for identifying the correct commutation relations for a Maxwell field \cite{LEADER2016303, Barnett_2016, Enk1994}. Here $\epsilon_{ijk}$ is the Levi-Civita tensor and $i,j,k=k_x,k_y,k_z$.

We verify the consistency of the commutation relations by numerically calculating the expectation values of the commutation relations and their respective quantum OAM counterparts in Table \ref{comm}.
\begin{table}[!htbp]
\centering
\begin{tabular}{|c|c|c|c|c|c|c|c|}
\cline{1-2} \cline{4-5} \cline{7-8}
$\langle [\hat{L}_x,\hat{L}_y] \rangle/{il\hbar}$& $\langle \hat{L}_z \rangle/l$  &&$\langle [\hat{L}_y,\hat{L}_z] \rangle/{il\hbar}$& $\langle \hat{L}_x \rangle/l$ &&$\langle [\hat{L}_z,\hat{L}_x] \rangle/{il\hbar}$& $\langle \hat{L}_y \rangle/l$ \\
\cline{1-2} \cline{4-5} \cline{7-8}
0.7071& 0.7071  &&1.069$\times 10^{-5}$& 0 &&0.7071& 0.7071 \\
\cline{1-2} \cline{4-5} \cline{7-8}
\end{tabular}
\caption{Comparison of the expectations of commutators of the quantum OAM operators against the expectation values of the OAM for validating the quantum commutation relation $[\hat{L}_i,\hat{L}_j]=i\hbar\epsilon_{ijk}\hat{L}_k$. The calculations are performed over a parameter space $n=1$, $\theta_c/\pi={0.05, 0.1, 0.2, 0.3}$ and $l={1, 10, 40, 80, 120, 160, 200}$. The standard deviations of the numerical calculations are under $10^{-5}$.}
\label{comm}
\end{table}

One can also derive these commutation relations analytically by employing the following bosonic commutation relations on Equation \ref{eq2}:
\begin{equation}
        [\hat{a}_{k,\lambda},\hat{a}^\dagger_{k',\lambda'}] = \delta_{\lambda,\lambda'}\delta^3(\bm{k-k'}), \quad [\hat{a}_{k,\lambda},\hat{a}_{k',\lambda'}]=[\hat{a}^\dagger_{k,\lambda},\hat{a}^\dagger_{k',\lambda'}]=0 \nonumber
\end{equation}

\subsection{Quantum OAM noise}

In this section, we derive the fluctuations or noise in the quantum OAM, a universal attribute present in all quantum pulses. The uncertainties inherent to the angular momentum stems from the Heisenberg uncertainties of the non-commuting quantum OAM operators (Appendix G). With the increase in the photon’s angular momentum the quantum fluctuations are pronounced, enhancing its detectability in experiments. The first-ever evidence of the presence of such fluctuations have been reported by \cite{Yang2021}, yet it fails to provide a complete 3D picture of the quantum fluctuations, as the noise was confined to the transverse plane. Thus, a universal 3D picture of the quantum OAM noise has not been developed till date. We find that the quantum spatiotemporal twisted pulses have fluctuations present along all three dimensions in $k-$space. In other words, the presence of a transverse OAM component directly leads to a non-vanishing longitudinal OAM noise.

In Appendix E, we explicitly derive the expressions for the OAM noises for the spatiotemporal pulse along the three directions in $k-$space, which directly originates from the variances of the quantum OAM operators:
\begin{equation}
    \Delta \hat{L}_i=\sqrt{\langle \hat{L}_i^2 \rangle - \langle \hat{L}_i \rangle ^2}
\end{equation}

\begin{figure}[t!]
    \centering
    \includegraphics[width=1\textwidth]{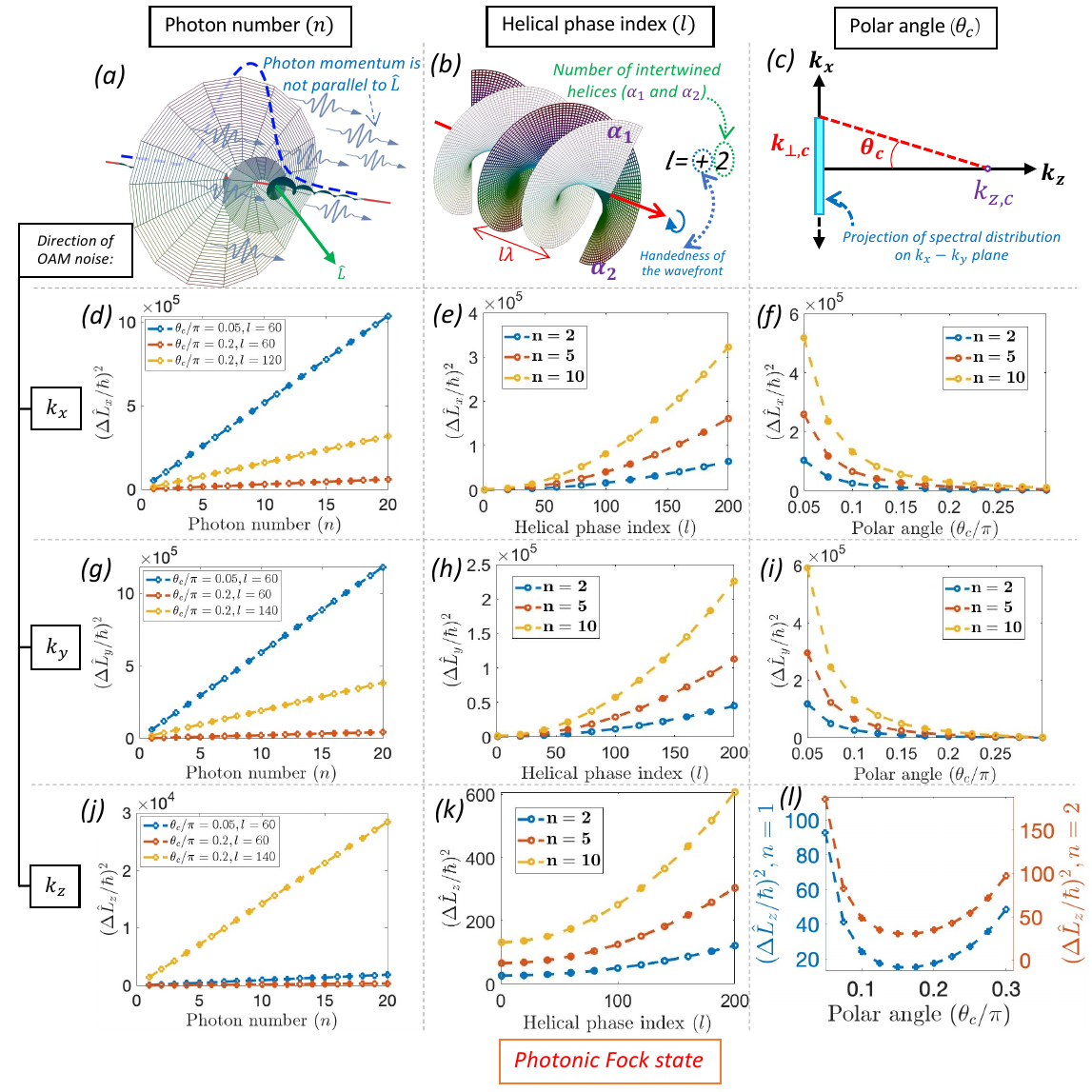}
    \caption{\textbf{OAM noise dependence on photon number ($n$), helical phase index ($l$) and polar angle ($\theta_c$) for the photonic Fock state pulse}. The pulse length is $\tau=2$ nm for center wave-length $\lambda_c=500$ nm for all the simulations in this paper. Other parameters: $\theta_c$ = $0.2\pi$ for the noise vs $l$ plots and $l$ = 60 for the noise vs $\theta_c$ plots.}
    \label{F2}
\end{figure}

\begin{figure}[t!]
    \centering
    \includegraphics[width=1\textwidth]{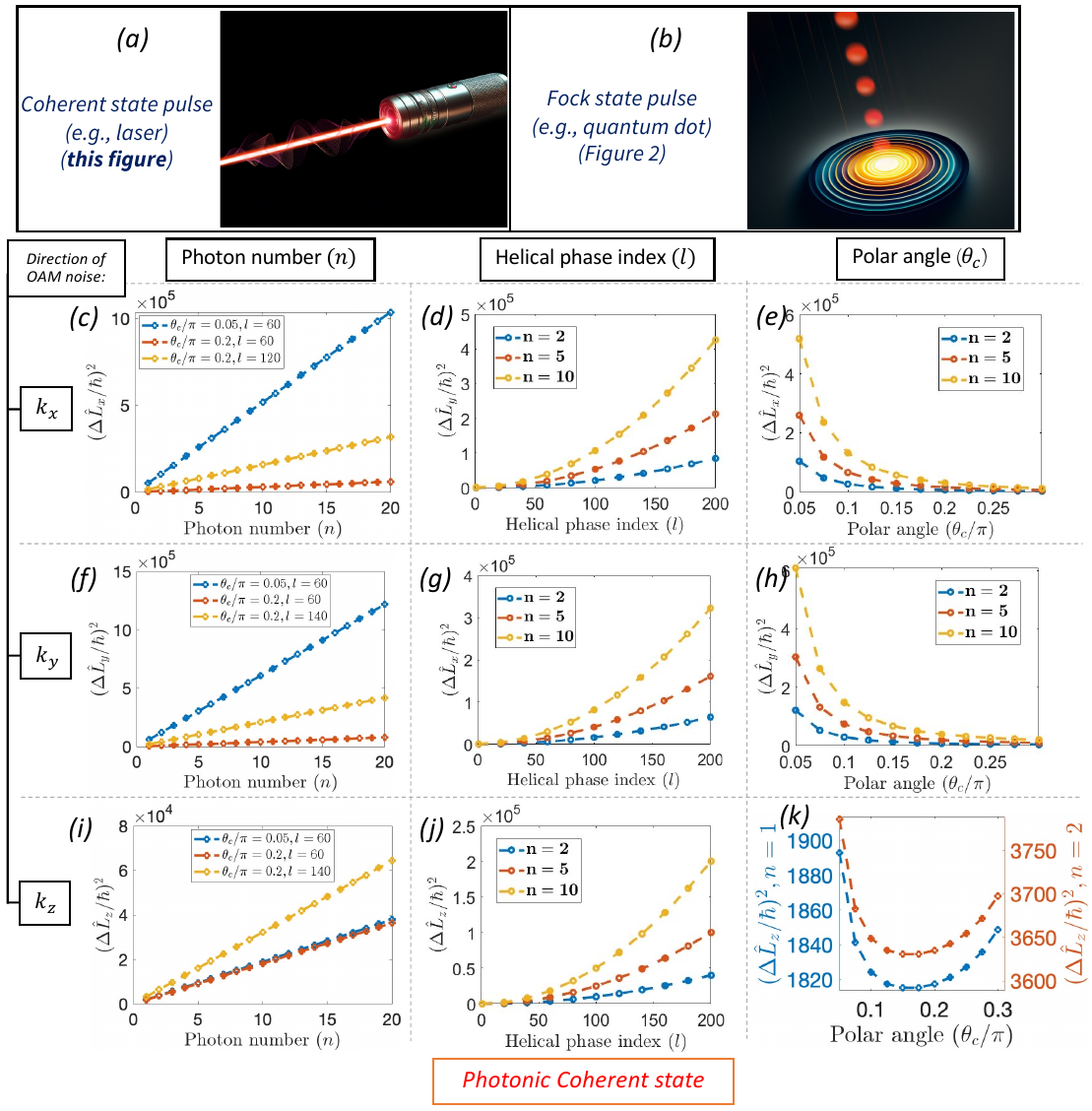}
    \caption{\textbf{OAM noise dependence on photon number ($n$), helical phase index ($l$) and polar angle ($\theta_c$) for the photonic coherent state pulse.} The parameters for the plots: $\theta_c$ = $0.2\pi$ for the noise vs $l$ plots and $l$ = 60 for the noise vs $\theta_c$ plots.}
    \label{F3}
\end{figure}

For the photonic Fock-state and coherent state pulses:
\begin{eqnarray}
    \label{eq4}
    \Delta \hat{L}_i^2 \Big|_{\text{Fock}} = \langle L_i^2 \rangle - \langle L_i \rangle^2 \Bigg|_{\text{Fock}}
     =& n\hbar^2 \Bigg[\int d^3k \xi_l^*(\bm{k}) (\boldsymbol{k\times \nabla_k})^2_i \xi_l(\bm{k}) \\ \nonumber
      & -\langle n_{\xi,\lambda}| \hat{\bm{L}}_i |n_{\xi,\lambda} \rangle ^2  \Bigg] ,
\end{eqnarray}
\begin{eqnarray}
    \label{eq5}
    \Delta \hat{L}_i^2 \Big|_{\text{coherent}} &= \langle L_i^2 \rangle - \langle L_i \rangle^2 \Bigg|_{\text{coherent}} \\ \nonumber
     &= \Tilde{n}\hbar^2 \Bigg[ \int d^3k \xi_l^*(\bm{k}) (\boldsymbol{k\times \nabla_k})^2_i \xi_l(\bm{k}) \Bigg] ,\quad i=k_x,k_y,k_z.
\end{eqnarray}

Although the vector potential, the conjugate momentum operators, and the bosonic ladder operators are polarization-dependent, we show in Appendix C that the quantum properties from the expectation values primarily stem from the SAF, which is polarization-independent. Thus the OAM noise given in equations \ref{eq4} and \ref{eq5} are independent of the polarization $\lambda$ of the pulse. Moreover, due to the absence of photon-photon interactions, this noise is purely a product of photon statistics.

In Fig. \ref{F2}, we have plotted the evolution of OAM noise as a function of the photon number $n$, helical phase index $l$ and the polar angle $\theta_c$ for a Fock state STOV pulse. Fig. \ref{F2}(a) marks an important signature of STOVs, where the momentum of the photon is not parallel to the orbital angular momentum. Figs. \ref{F2}(d,g,j) show a linear dependency of the square of the OAM noise on the photon number, resulting in high fluctuations with increasing photon counts. From Eq. \ref{eq1}, one can deduce that the helical phase index has a polynomial dependence on the square of the OAM noise. This signature can also be seen in the plots \ref{F2}(e,h,k).

We obtain a unique feature of the OAM noise in the longitudinal direction which is absent in the transverse plane. From Figs. \ref{F2}(f,i), we observe that the quantum OAM noise decays rapidly with the increase in the polar angle. However, the longitudinal OAM noise in Fig. \ref{F2}(l) experiences a growth as the polar angle approaches the paraxial limit. This result is rather expected, since suppressing the transverse noise shall increase the uncertainty in the longitudinal OAM in accordance to the Heisenberg uncertainty rule.

The dependence of the OAM noise for the photonic coherent state in Fig. \ref{F3} shows similar signatures as the photonic Fock-state. Yet, the underlining difference lies in the fact that the OAM noise for coherent state pulses exceeds that of the Fock-state pulses. It can also be seen as a corollary from the two equations \ref{eq4} and \ref{eq5}. This signature is prominent in Figs. \ref{F3}(j,k), where the helical phase index dependence plot shows an increment of the square of OAM fluctuations by three orders in magnitude and polar angle dependence plots reflect an increment of an order in magnitude, while compared to the Fock state counterparts in Figs. \ref{F2}(k,l). These signatures are exclusive for the longitudinal OAM noise, further strengthening our claims for experimental validation with the existing setups \cite{Ji2020,Suprano2021,Behera2020,Aboushelbaya2019,Lavery2013,Xie2018,Fatkhiev2021}, especially with a photonic coherent state pulse.


\subsection{Temporal evolution of OAM Noise Density}

In this section, we show the spatial distribution the OAM fluctuation, and its evolution over time. We define this quantity as the OAM noise density. Similar to the case of the OAM noise, the OAM noise density purely exists in the quantum domain.

The momentum density operator for the OAM is given by
\begin{align}
    \hat{\mathcal{L}_i} = & \epsilon_0 \hat{E}_\perp^j(r,t)(\bm{r}\times\boldsymbol{\nabla})_i\hat{A}_\perp^j(r,t), \qquad \qquad i= k_x,k_y,k_z \\ \nonumber
    = &\frac{i\hbar}{2(2\pi)^{3}}\int d^{3}k\int d^{3}k'\sum_{\lambda,\lambda'=\pm1}\sqrt{\frac{\omega(\bm{k})}{\omega'(\bm{k'})}}\left[\hat{a}_{k,\l}\bm{\epsilon}^{j}(\bm{k},\l)e^{i(\bm{k.r}-\omega t)}-h.c.\right] \\ \nonumber
    &\qquad \left(\bm{r\times\nabla}\right)_i\left[\hat{a}_{k',\l'}\bm{\epsilon}^{j}(\bm{k'},\l')e^{i(\bm{k'}.\bm{r}-\omega't)}+h.c.\right],
    \label{eq7}
\end{align}
\noindent where $\bm{\epsilon}^j(k,\l)$ are the circular polarization vectors ($\l=-1$ for left-handed polarization and $\l=+1$ for right-handed polarization). $\hat{A}_\perp$ and $\hat{E}_\perp$ are the operators for the vector potential and the electric field respectively, in the transverse plane. Although this framework can be generalized beyond the paraxial regime, for simplicity, we consider the paraxial limit, where the polar angle $\theta_c \rightarrow 0$. Under this assumption, we obtain $\omega(k)\approx \omega(k')$, the details of which is presented in Appendix F.

We primarily focus on mapping the noise density along the $z-$direction, as it is unique to the STOV and its potential for immediate experimental validation. For the \textit{n}-photon Fock-state pulse, the quantum OAM noise density in the longitudinal direction is given by (Appendix F):
\begin{equation}
    \Delta\hat{\mathcal{L}}_z = \sqrt{\langle n_{\xi,\l}|\hat{\mathcal{L}_{z}}^{2}(r,t)|n_{\xi,\l}\rangle - \langle n_{\xi,\l}|\hat{\mathcal{L}_{z}}(r,t)|n_{\xi,\l}\rangle^{2}}
\end{equation}

\begin{figure}[!ht]
    \centering
    \includegraphics[width=1\textwidth]{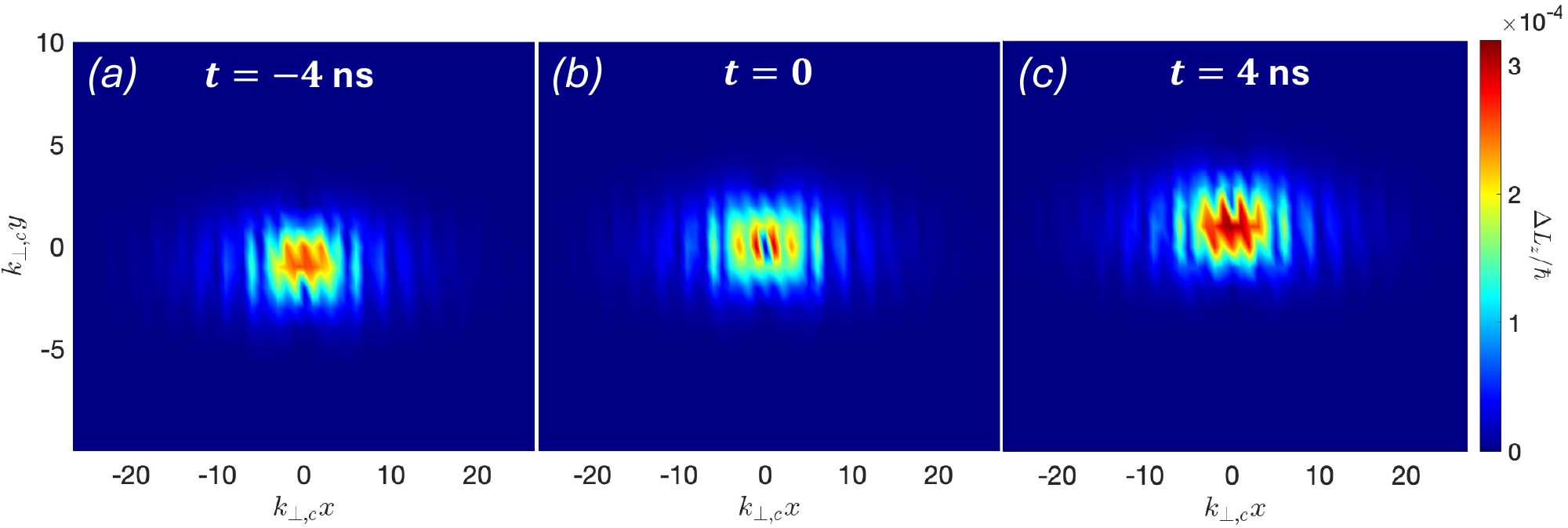}
    \caption{\textbf{OAM noise density in the longitudinal direction} in the paraxial limit for the right-handed STOV. Figures (a) $\rightarrow$ (c) show the evolution of the OAM noise density at different times \textit{t}, in the frame $z=ct$, for center wavelength $\lambda_c=500$ nm and pulse length $\tau=2ns$. We used dimensionless variables $k_{\perp,c}x$ and $k_{\perp,c}y$ for the plots. Here, $n=10$, $\theta_c=0.1$ and $l=2$. Higher $l$ and working with coherent state pulses shall increase this noise by at least two orders in magnitude.}
    \label{F5}
\end{figure}

Here, $\langle n_{\xi,\l}|\hat{\mathcal{L}_{z}}(r,t)|n_{\xi,\l}\rangle$ is the mean OAM density of the STOV and $\langle n_{\xi,\l}|\hat{\mathcal{L}_{z}}^2(r,t)|n_{\xi,\l}\rangle$ originates from the two-point correlation of the OAM density $\langle n_{\xi,\l}|\hat{\mathcal{L}_{z}}(r,t)\hat{\mathcal{L}_{z}}(r',t')|n_{\xi,\l}\rangle$ in the limit $r-r'\rightarrow 0$. Figure \ref{F5} shows the texture of the OAM fluctuations along the direction of propagation in real space. The full 3D picture of the OAM fluctuations for any arbitrary quantum pulse can be computed using similar calculations as in Appendix F.

In Fig. \ref{F5}, we have plotted the temporal evolution of the OAM noise density. The longitudinal OAM density fluctuations vanishes in the central region at time $t=0$, which corresponds to the phase singularity for the STOV \cite{Bliokh2012}. In other words, similar to the energy density, the OAM noise density of the STOV propagates with a moving smoke-ring-like signature. Although we have used $l=2$ for the plots, but as highlighted in the previous sections, working with a coherent state quantum pulse with higher values of $l$ enhances the OAM noise by two orders of magnitude or more, further making it amendable for experimental validation (Figs. \ref{F2}, \ref{F3}).

\section{Discussions}
The presence of large quantum OAM fluctuations and its spatial distribution in real space, especially in the longitudinal direction can be readily validated with experiments. These quantum fluctuations explicitly depend on the physical quantities (photon number, helical phase index and polar angle of the beam), which can be tuned and measured in an experimental environment. Some of the experiments done in the past few years such as the photocurrent detection of the orbital angular momentum of light \cite{Ji2020}, OAM coupling in elastic photon-phonon scattering \cite{Aboushelbaya2019}, and optomechanical OAM detection through optically induced torque \cite{Strasser2022, He2016, Padgett2011,Behera2020,Lavery2013} can be used as promising approaches to show the experimental signatures of the OAM noise. Furthermore, the texture of these large OAM noise (OAM noise density) can be probed by investigating the OAM coupling with quantum sensors such as the nitrogen-vacancy centers in diamonds \cite{Kalhor2021}.

In summary, our work establishes a robust framework for understanding the quantum nature of a general spatiotemporal twisted pulse. Beyond classical and semi-classical approaches, our framework includes both transverse and longitudinal components of Orbital Angular Momentum (OAM), shedding light on the pure quantum phenomena of fluctuations in the OAM. Our findings emphasize the need for further experimental probes to harness the predicted quantum properties of STOVs. The insights gained from this work lays the groundwork for leveraging STOVs in emerging quantum technologies, pushing the boundaries of information processing and optical manipulation at the quantum level.

\section{Acknowledgements}
This work is supported by the funding from Army Research Office (W911NF-21-1-0287).

\section{Data Availability Statement}
All data that support the findings of this study are included within the article (and any supplementary files).

\appendix

\section{Photon AM quantum operators}

The plane-wave expansion of the electromagnetic fields in free space have the forms:
\begin{align*}
\hat{A}^{j}_\perp & =\int d^{3}k\sum_{\lambda=\pm 1}\sqrt{\frac{\hbar}{2\varepsilon_{0}\omega_{k}(2\pi)^{3}}}[\hat{a}_{k,\lambda}\epsilon^{j}(\boldsymbol{k},\lambda)e^{i(\boldsymbol{k.}\textbf{r}-\omega_k t)}+h.c.],\\
\hat{E}^{j}_\perp & =i\int d^{3}k\sum_{\lambda=\pm 1}\sqrt{\frac{\hbar\omega_{k}}{2\varepsilon_{0}(2\pi)^{3}}}[\hat{a}_{k,\lambda}\epsilon^{j}(\boldsymbol{k},\lambda)e^{i(\boldsymbol{k.}\textbf{r}-\omega_k t)}-h.c.].
\end{align*}

Here, we have used the circular polarization vectors ${e}(\bm{k},\lambda = \pm 1)$, corresponding to the two observable degrees of freedom of the photon (refer \cite{Yang2022}). One can use the bosonic commutation relations
\begin{align}
[\hat{a}_{k,\lambda},\hat{a}_{k',\lambda'}^{\dagger}] &=\delta_{\lambda,\lambda'}\delta^{3}(\boldsymbol{k-k}');\\ \nonumber
[\hat{a}_{k,\lambda},\hat{a}_{k',\lambda'}] &=[\hat{a}_{k,\lambda}^{\dagger},\hat{a}_{k',\lambda'}^{\dagger}]=0,
\end{align}

\noindent to derive the following relations for the vector potential and the conjugate momentum (orthonormality condition of the polarization vectors: $\sum_{\lambda=\pm1} e^j(\bm{k},\lambda)e^i(\bm{k},\lambda)=\delta_{\perp}^{ij}$):
\begin{align*}
[\hat{A}^{i}_\perp(\bm{r},t),\hat{\pi}^{j}_\perp(\bm{r-r}',t)] & =\frac{i\hbar}{\varepsilon_0}\delta_{\perp}^{ij}\delta^{3}(\bm{r-r}'),\\{}
[\hat{A}^{i}_\perp(\bm{r},t),\hat{A}^{j}_\perp(\bm{r}',t)] & =[\hat{\pi}^{i}_\perp(\bm{r},t),\hat{\pi}^{j}_\perp(\bm{r}',t)]=0.
\end{align*}

We can derive the observable orbital angular momentum operator using Noether's theorem and eliminating the gauge-dependent variables

\begin{equation}
    \hat{\bm{L}} =-\int d^{3}r[\hat{\pi}_\perp^{j}(\bm{r\times\nabla})\hat{A}_{\perp}^j]=-i\hbar \int d^3k \sum_{\lambda=\pm 1}[\hat{a}_{k,\lambda}^\dagger(\boldsymbol{k \times \nabla_k})\hat{a}_{k,\lambda}].
    \label{OAM_obs}
\end{equation}
In cylindrical coordinates,

\begin{eqnarray}
    (\boldsymbol{k\times\nabla_{k}})_{x} & = & \Bigg(\rho_{k}\sin\phi_{k}\frac{\partial}{\partial k_{z}}-k_{z}\sin\phi_{k}\frac{\partial}{\partial\rho_{k}}-\frac{k_{z}}{\rho_{k}}\cos\phi_{k}\frac{\partial}{\partial\phi_{k}}\Bigg),\\
    (\boldsymbol{k\times\nabla_{k}})_{y} & = & -\Bigg(\rho_{k}\cos\phi_{k}\frac{\partial}{\partial k_{z}}-k_{z}\cos\phi_{k}\frac{\partial}{\partial\rho_{k}}+\frac{k_{z}}{\rho_{k}}\sin\phi_{k}\frac{\partial}{\partial\phi_{k}}\Bigg),\\
    (\boldsymbol{k\times\nabla_{k}})_{z} & = & \frac{\partial}{\partial\phi_{k}}.
\end{eqnarray}

\section{SAF of the Spatiotemporal Optical Vortex}

The single-photon wave-packet creation operator for any twisted pulse is given by:
\begin{equation}
    \hat{a}^\dagger_{\xi,\lambda}=\int d^3k \xi_l(\bm{k}) \hat{a}^\dagger_{k,\lambda}. \nonumber
\end{equation}
This operator is a superposition of all the plane waves with amplitude function $\xi_l$ having a common fixed polarization $\lambda$. For this creation (and annihilation) operators to follow the bosonic commutation relations, it is mandatory for the amplitude function to be normalizable.

The amplitude function for the monochromatic twisted Bessel-Gaussian pulse can be expressed as \cite{Yang2021}:
\begin{eqnarray*}
\e_{l}(\r,k_{z},\o) & = & \frac{1}{\sqrt{2\pi}}\Big(\frac{2\s_{z}^{2}}{\pi}\Big)^{\frac{1}{4}}\Big(\frac{2\sigma_{\rho}^{2}}{\pi k_{\perp,c}^{2}}\Big)^{\frac{1}{4}}\exp\Bigg[-\s_{\rho}^{2}(\r-k_{\perp,c})^{2}\Bigg]\\ \nonumber
 &  & \times\exp\Bigg[-\s_{z}^{2}(k_{z}-k_{z,c})^{2}+il\o\Bigg]
\end{eqnarray*}
Here, the OAM is along the $k_z$ direction, and forms a cone in the $k$-space. We can get a component of the OAM in the transverse direction by rotating the spectral distribution along $k_x$ with $k_{z,c}$ at the center using the following transformations:
\begin{eqnarray}
    k_{y} \rightarrow k_{y}'=k_{y}\cos\theta+(k_{z}-k_{z,c})\sin\theta,\\
    k_{z} \rightarrow k_{z}'=-k_{y}\sin\theta+(k_{z}-k_{z,c})\cos\theta+k_{z,c},\\
    \phi_k = \tan^{-1}\Big(\frac{k_y}{k_x}\Big) \rightarrow \phi_k' = \tan^{-1}\frac{\r\sin\o\cos\theta+(k_{z}-k_{z,c})\sin\theta}{\r\cos\o},
\end{eqnarray}
and the SAF has the form:
\begin{eqnarray}
    &\e_{l}(\r,k_{z},\o)\\ \nonumber
    &=  \frac{1}{\sqrt{2\pi}}\Big(\frac{2\s_{z}^{2}}{\pi}\Big)^{\frac{1}{4}}\Big(\frac{2\sigma_{\rho}^{2}}{\pi k_{\perp,c}^{2}}\Big)^{\frac{1}{4}}\\ \nonumber
     &\times\exp\Bigg[-\s_{\rho}^{2}(\sqrt{(\r\cos\o)^{2}+(\r\sin\o\cos\theta+(k_{z}-k_{z,c})\sin\theta)^{2}}-k_{\perp,c})^{2}\Bigg]\\ \nonumber
    & \times\exp\Bigg[-\s_{z}^{2}(-\r\sin\o\sin\theta+(k_{z}-k_{z,c})\cos\theta)^{2}\\ \nonumber
    & \qquad+il\tan^{-1}\frac{\r\sin\o\cos\theta+(k_{z}-k_{z,c})\sin\theta}{\r\cos\o}\Bigg]
\end{eqnarray}

Before we proceed, we verify that the photon wavefunction is a solution of the general wave equation, without necessitating the paraxial approximation. In real space, we can always define the wavefunction of the STOV as:
\begin{equation}
    \psi_{l}(\bm{r},t) = \xi_{l}(\bm{r})e^{i(\bm{\bm{k.r}}-\omega t)}.
\end{equation}

Fourier transform of the Laplacian of the wavefunction (assuming $r^{2}\nabla\p$
goes to zero as $r\rightarrow\infty$):
\begin{eqnarray*}
    \iint_{-\text{\ensuremath{\infty}}}^{\infty}d\bm{r'}dt'\nabla^{2}\text{\ensuremath{\p}}_{l}(\bm{r'},t)e^{i(\bm{\textbf{k.r}'-\omega}t'\bm{)}} & = & -\iint_{-\text{\ensuremath{\infty}}}^{\infty}d\bm{r'}\nabla\e_{l}(\bm{r})\nabla e^{i(\bm{k.(r+r')}-\omega(t-t'))}\\
     & = & \iint_{-\text{\ensuremath{\infty}}}^{\infty}d\bm{r'}\e_{l}(\bm{r})\nabla^{2}e^{i(\bm{\textbf{k.r}'}-\omega t')}\\
     & = & -|\bm{k}|^{2}\iint_{-\text{\ensuremath{\infty}}}^{\infty}d\bm{r'}\text{\ensuremath{\p}}_{l}(\bm{r'},t)e^{i(\bm{\textbf{k.r}'}-\omega t')}\\
     & = & -|\bm{k}|^{2}\text{\ensuremath{\p}}_{l}(\bm{k,}\omega)
\end{eqnarray*}

\begin{eqnarray*}
\iint_{-\text{\ensuremath{\infty}}}^{\infty}d\bm{r'}dt'\text{\ensuremath{\partial}}_{t'}\text{\ensuremath{\p}}_{l}(\bm{r},t') & = & -\omega^{2}\iint_{-\text{\ensuremath{\infty}}}^{\infty}d\bm{r'}dt'\e_{l}(\bm{r}')e^{i(\bm{k.(r+r')}-\omega(t-t'))}\\
 & = & -\omega^{2}\text{\ensuremath{\p}}_{l}(\bm{k,}\omega)
\end{eqnarray*}
The wave equation $\nabla^2 \psi_l = \partial^2_t \psi_l$ thus becomes:
\begin{equation}
    (\omega^{2}-|\bm{k}|^{2})\text{\ensuremath{\p}}_{l}(\bm{k,}\omega)  =  0.
\end{equation}
Here, $|\bm{k}|$ is not a constant value (Fig. \ref{AF1}), which makes the frequency $\omega$ dependent on $\bm{k}$. Also, this relation tells us that in the $k^\mu$-space, there are only three degrees of freedom ($k^0 = \omega = |\bm{k}|$), thus implying $\text{\ensuremath{\p}}_{l}(\bm{k,}\omega) = \e_l (\bm{k})$.

\section{Photonic Fock State and Coherent State pulses}

The wavefunction of the $n$-photon Fock state twisted pulse can be written as:
\begin{equation}
    |n_{\xi,\lambda} \rangle= \frac{1}{\sqrt{n!}}\Big(\hat{a}_{\xi,\lambda}^\dagger \Big)^n |0\rangle.
\end{equation}
We can prove the orthonormal condition $\langle n_{\xi,\lambda} | n'_{\xi,\lambda} \rangle = \delta _{nn'}$ using the bosonic commutation relation $\Big[\hat{a}_{\xi,\lambda},\hat{a}^\dagger_{\xi,\lambda}]=1$.

One can easily show that this wavefunction satisfies the following relations:
\begin{eqnarray}
    &\hat{a}_{k,\lambda'}| n_{\xi,\lambda} \rangle  = \sqrt{n}\xi_l(\bm{k}) \delta_{\lambda,\lambda'}|(n-1)_{\xi,\lambda} \rangle, \\ \nonumber
    &\hat{a}_{k,\lambda'}^\dagger| n_{\xi,\lambda} \rangle  = \sqrt{n+1}\xi_l(\bm{k}) \delta_{\lambda,\lambda'}|(n+1)_{\xi,\lambda} \rangle, \\ \nonumber
    &\langle n_{\xi,\lambda} | \int d^3k \sum_{\lambda'=\pm 1} \hat{a}_{k,\lambda'}^\dagger \hat{a}_{k,\lambda'} | n_{\xi,\lambda} \rangle = n.
\end{eqnarray}
Similarly, a mean-$\Tilde{n}$ photon Coherent state pulse can be constructed as ($|\alpha|^2 = \Tilde{n}$):
\begin{equation}
    |\alpha_{\xi,\lambda} \rangle = e^{-\Tilde{n}/2}\sum_{k=0}^\infty \frac{\alpha^k}{\sqrt{k!}}|n_{\xi,\lambda}\rangle.
\end{equation}
Even for this state, one can easily show the following relations:
\begin{eqnarray}
    &\hat{a}_{k,\lambda'}| \alpha_{\xi,\lambda} \rangle  = \alpha \xi_l(\bm{k}) \delta_{\lambda,\lambda'}|\alpha_{\xi,\lambda} \rangle, \\ \nonumber
    &\hat{a}_{k,\lambda'}^\dagger| \alpha_{\xi,\lambda} \rangle  = \alpha \xi_l(\bm{k}) \delta_{\lambda,\lambda'}|\alpha_{\xi,\lambda} \rangle, \\ \nonumber
    &\langle \alpha_{\xi,\lambda} | \int d^3k \sum_{\lambda'=\pm 1} \hat{a}_{k,\lambda'}^\dagger \hat{a}_{k,\lambda'} | \alpha_{\xi,\lambda} \rangle = \alpha^*\alpha = \Tilde{n}.
\end{eqnarray}

\section{Expectation values of the quantum OAM operator}
\begin{table}[ht]
\centering
\begin{tabular}{|c|c|c|c|c|c|c|}
\cline{2-7}
\multicolumn{1}{c|}{}& \multicolumn{2}{c|}{$\langle \hat{L}_x \rangle/\hbar$} & \multicolumn{2}{c|}{$\langle \hat{L}_y \rangle/\hbar$} & \multicolumn{2}{c|}{$\langle \hat{L}_z \rangle/\hbar$} \\
\cline{2-7}
\multicolumn{1}{c|}{}& mean &std. dev. & mean &std. dev. & mean &std. dev. \\ \hline
Calculated value & 6.9889e-05 & 8$\times 10^{-5}$ & 14.1422 & 1$\times 10^{-4}$ & 14.1422 & 9$\times 10^{-5}$ \\ \hline
Actual value & \multicolumn{2}{c|}{0} & \multicolumn{2}{c|}{14.1422} & \multicolumn{2}{c|}{14.1422} \\ \hline
\end{tabular}
\caption{OAM for the quantum Bessel-Gaussian spatiotemporal pulse along $k_x,\;k_y$ and $k_z$, for $n\:(\Tilde{n})=2,\;l=10$ (computed over $0.05\pi \leq \theta_c \leq 0.3\pi$)}
\end{table}

In Table D1, we have provided the Orbital Angular Momentum (OAM) values corresponding to the quantum spatiotemporal pulse along the three directions in $k-$space, applicable to both the Fock-state and coherent state pulses. This is equivalent to the projection of the established semiclassical mean OAM ($nl\hbar$) along the $k_x,k_y,k_z$ directions, affirming the consistency of this theory with the literature.

\section{OAM fluctuations of the spatiotemporal optical vortex}

The square of the quantum OAM operator is given by:
\begin{eqnarray*}
\hat{L}_i^2 = &-\hbar^2 \Bigg[ \int d^3k \int d^3k' \sum_{\lambda,
\lambda'=\pm 1} \hat{a}_{k,\lambda}^\dagger \hat{a}_{k',\lambda'}^\dagger(\boldsymbol{k \times \nabla_k})_i(\boldsymbol{k' \times \nabla_{k'}})_i\hat{a}_{k,\lambda} \hat{a}_{k',\lambda'} \Bigg] \\ 
&- \hbar^2 \int d^3k \sum_{\lambda=\pm 1}\hat{a}_{k,\lambda}^\dagger(\boldsymbol{k \times \nabla_k})_i^2\hat{a}_{k,\lambda},
\end{eqnarray*}
where $i=k_x,k_y,k_z$. For a photonic Fock-state pulse,
\begin{eqnarray*}
    \langle n_{\xi,\lambda}| \hat{L}_i^2 |n_{\xi,\lambda} \rangle &= \hbar^2 \Bigg[ & n(n-1) \Bigg( \int d^3k \xi_l^*(\bm{k}) (\boldsymbol{k\times \nabla_k})_i \xi_l(\bm{k}) \Bigg)^2 \\ \nonumber
    & & + n \int d^3k \xi_l^*(\bm{k}) (\boldsymbol{k\times \nabla_k})^2_i \xi_l(\bm{k}) \Bigg] \\ \nonumber
    &=\hbar^2 \Bigg[& n(n-1) \langle n_{\xi,\lambda}| \hat{\bm{L}}_i |n_{\xi,\lambda} \rangle ^2 + n \int d^3k \xi_l^*(\bm{k}) (\boldsymbol{k\times \nabla_k})^2_i \xi_l(\bm{k}) \Bigg], \nonumber
\end{eqnarray*}
and for a photonic coherent-state pulse,
\begin{eqnarray*}
    \langle \alpha_{\xi,\lambda}| \hat{L}_i^2 |\alpha_{\xi,\lambda} \rangle &= \hbar^2 \Bigg[ & \Tilde{n}^2 \Bigg( \int d^3k \xi_l^*(\bm{k}) (\boldsymbol{k\times \nabla_k})_i \xi_l(\bm{k}) \Bigg)^2 \\ \nonumber
    & &+ \Tilde{n} \int d^3k \xi_l^*(\bm{k}) (\boldsymbol{k\times \nabla_k})^2_i \xi_l(\bm{k}) \Bigg]\\\nonumber
    &=\hbar^2 \Bigg[& \Tilde{n}^2 \langle n_{\xi,\lambda}| \hat{\bm{L}}_i |\alpha_{\xi,\lambda} \rangle ^2 + \Tilde{n} \int d^3k \xi_l^*(\bm{k}) (\boldsymbol{k\times \nabla_k})^2_i \xi_l(\bm{k}) \Bigg]. \nonumber
\end{eqnarray*}
Therefore, we can express the OAM noise for the photonic Fock and coherent states as
\begin{eqnarray}
    \Delta \hat{L}_i^2 \Big|_{\text{Fock}} = \langle \hat{L}_i^2 \rangle - \langle \hat{L}_i \rangle^2 \Bigg|_{\text{Fock}}
     =& n\hbar^2 \Bigg[\int d^3k \xi_l^*(\bm{k}) (\boldsymbol{k\times \nabla_k})^2_i \xi_l(\bm{k}) \\ \nonumber
      & -\langle n_{\xi,\lambda}| \hat{\bm{L}}_i |n_{\xi,\lambda} \rangle ^2  \Bigg],
\end{eqnarray}
\begin{eqnarray}
    \Delta \hat{L}_i^2 \Big|_{\text{coherent}} &= \langle \hat{L}_i^2 \rangle - \langle \hat{L}_i \rangle^2 \Bigg|_{\text{coherent}} \\ \nonumber
     &= \Tilde{n}\hbar^2 \Bigg[ \int d^3k \xi_l^*(\bm{k}) (\boldsymbol{k\times \nabla_k})^2_i \xi_l(\bm{k}) \Bigg].
\end{eqnarray}

\section{OAM noise density}
The OAM density operator is given by:
\begin{equation}
    \bm{\hat{\mathcal{L}}}=\epsilon_{0}\hat{E}_{\perp}^{j}(r,t)(\bm{r\times\nabla})\hat{A}_{\perp}^{j}(r,t).
\end{equation}
where $\hat{E}_{\perp}^{j}(r,t)$ and $\hat{A}_{\perp}^{j}(r,t)$ are the electric and vector potential operators on the transverse plane. The component along the longitudinal direction is not an observable quantity, since it is purely gauge-dependent \cite{Yang2022}. The field operators can be written as:
\begin{align}
    \hat{E}_{\perp}^{j}(r,t)&=i\int d^{3}k\sum_{\l=\pm1}\sqrt{\frac{\hbar\omega}{2\epsilon_{0}(2\pi)^{3}}}\left[\hat{a}_{k,\l}\bm{\epsilon}^{j}(k,\l)e^{i(\textbf{k.r}-\omega t)}-h.c.\right].\\
    \hat{A}_{\perp}^{j}(r,t)&=\int d^{3}k'\sum_{\l=\pm1}\sqrt{\frac{\hbar}{2\omega'\epsilon_{0}(2\pi)^{3}}}\left[\hat{a}_{k',\l}\bm{\epsilon}^{j}(k',\l)e^{i(\bm{k}'.\textbf{r}-\omega't)}+h.c.\right].
\end{align}

Applying the paraxial approximation $\theta_c\rightarrow0$, we obtain:
\begin{align*}
    \omega	&=k_{z}\sqrt{1+\bigg(\frac{k_{x}^{2}+k_{y}^{2}}{k_{z}^{2}}\bigg)}\\
	&\approx k_{z}\Bigg(1+\frac{3}{4}\tan^{2}\theta_{c}\Bigg),\\
    \sqrt{\omega}	&\approx\sqrt{k_{z}}\Bigg(1+\frac{3}{8}\tan^{2}\theta_{c}\Bigg),\\
    \omega^{-1}&\approx\sqrt{k_{z}}\Bigg(1-\frac{3}{8}\tan^{2}\theta_{c}\Bigg),
\end{align*}
\noindent and $\sqrt{\omega(k)/\omega(k')}\approx1$. We can expand the $k$-dependent circular polarization vectors in a fixed lab frame as:
\begin{equation}
    \bm{\epsilon}(k,\lambda)=e^{-i\l\p_{k}}\cos^{2}(\frac{\theta_{c}}{2})e_{\l}-e^{i\l\p_{k}}\sin^{2}(\frac{\theta_{c}}{2})e_{-\l}-\frac{1}{\sqrt{2}}\sin(\theta_{c})e_{z} ,\nonumber
\end{equation}
\noindent where $e_{\l}=(e_{x}+i\l e_{y})/\sqrt{2}$. If we denote the wave-vector-dependent frame with $k'$, the inner product of the two polarization vectors under paraxial approximation thus become:
\begin{align*}
    \bm{\epsilon}^{j}(k,\l).\bm{e}^{j*}(k',\l)&=e^{-i\l(\phi_{k}-\phi_{k}')}\cos^{4}(\frac{\theta_{c}}{2})+e^{i\l(\phi_{k}-\phi_{k}')}\sin^{4}(\frac{\theta_{c}}{2})+\frac{\sin^{2}(\theta_{c})}{2}\\
    &\approx e^{-i\l(\p_{k}-\p_{k}')}.
\end{align*}
Hence, the OAM noise density operator along $z$ can be written as:
\begin{align}
    \hat{\mathcal{L}}_{z}&=\frac{i\hbar}{2(2\pi)^{3}}\int d^{3}k\int d^{3}k'\sum_{\l,\l'=\pm1}\left[\hat{a}_{k,\l}\bm{\epsilon}^{j}(k,\l)e^{i(\textbf{k.r}-\omega t)}-h.c.\right]\\ \nonumber
    &\qquad(x\frac{\partial}{\partial y}-y\frac{\partial}{\partial x})\left[\hat{a}_{k',\l'}\bm{\epsilon}^{j}(k',\l')e^{i(\bm{k}'.\textbf{r}-\omega't)}+h.c.\right]\\ \nonumber
    &=\frac{\hbar}{(2\pi)^{3}}\int d^{3}k\int d^{3}k'\sum_{\l=\pm1}(xk'_{y}-yk'_{x})\Bigg[\hat{a}_{k}^{\dagger}\hat{a}_{k'}e^{-i\l(\phi_{k}-\phi_{k}')}e^{-i((\bm{k}-\bm{k}').\textbf{r}-(\omega-\omega')t)}\Bigg]
\end{align}
Without loss of generality, we derive the OAM noise density for the \textit{n}-photon Fock-state pulse. The expectation value of the OAM density is:
\begin{equation}
    \langle \hat{\mathcal{L}}_{z} \rangle = \frac{n\hbar}{(2\pi)^{3}}\int d^{3}k\int d^{3}k'\sum_{\l=\pm1}(xk'_{y}-yk'_{x})\Bigg[\xi_l^*(k)\xi_l(k')e^{-i\l(\phi_{k}-\phi_{k}')}e^{-i((\bm{k}-\bm{k}').\textbf{r}-(\omega-\omega')t)}\Bigg].
\end{equation}
We define the effective wavefunction of the STOV in real space as
\begin{equation}
    \psi^l_\l (r,t)=\frac{1}{(2\pi)^{3}}\int d^3k \xi_l(k)e^{i(\textbf{k.r}-\omega t - \l \phi_k)}.
\end{equation}
Thus, the expectation value of the OAM density is:
\begin{equation}
    \langle \hat{\mathcal{L}}_{z} \rangle = \frac{n\hbar}{\sqrt{(2\pi)^{3}}}\sum_{\l=\pm1}\psi^{l*}_\l(r,t)\int d^{3}k(xk_{y}-yk_{x})\xi_l(k)e^{i(\textbf{k.r}-\omega t - \l \phi_k)}.
\end{equation}

For computing the fluctuations in the longitudinal OAM density, we start with calculating the two-point correlation of the OAM density:
\begin{align*}
    &\langle n_{\xi,\l}|\hat{\mathcal{L}_{z}}(r,t)\hat{\mathcal{L}_{z}}(r',t')|n_{\xi,\l}\rangle\\ &=\Bigg(\frac{\hbar}{2(2\pi)^{3}}\Bigg)^{2}\langle n_{\xi,\l}|\sum_{\l=\pm1}\int d^{3}k\int d^{3}k'(xk'_{y}-yk'_{x})\left[\hat{a}_{k,\l}\bm{\epsilon}^{j}(k,\l)e^{i(\textbf{k.r}-\omega t)}-h.c.\right]\\
    &\quad \left[\hat{a}_{k',\l'}\bm{\epsilon}^{j}(k',\l)e^{i(\bm{k}'.\textbf{r}-\omega't)}-h.c.\right]\int d^{3}k''\int d^{3}k'''(xk'''_{y}-yk'''_{x})\left[\hat{a}_{k'',\l}\bm{\epsilon}^{j}(k'',\l)e^{i(k''.r'-\omega''t')}-h.c.\right]\\
    &=2\Bigg(\frac{\hbar}{2(2\pi)^{3}}\Bigg)^{2}\langle n_{\xi,\l}|\sum_{\l=\pm1}\int d^{3}k\int d^{3}k'\int d^{3}k''\int d^{3}k'''(xk'_{y}-yk'_{x})(xk'''_{y}-yk'''_{x})\\
    &\quad \Bigg[\hat{a}_{k,\l}^{\dagger}e^{-i(\textbf{k.r}-\omega t-\l\p_{k})}\hat{a}_{k',\l}^{\dagger}e^{-i(\bm{k}'.\textbf{r}-\omega't-\l\p'_{k})}\hat{a}_{k'',\l}e^{i(k''.r'-\omega''t'-\l\p'_{k})}\hat{a}_{k''',\l}e^{i(k'''.r'-\omega'''t'-\l\p'''_{k})}\\
    &\quad+\hat{a}_{k,\l}^{\dagger}e^{-i(\textbf{k.r}-\omega t-\l\p_{k})}\hat{a}_{k',\l}e^{i(\bm{k}'.\textbf{r}-\omega't-\l\p'_{k})}\hat{a}_{k'',\l}^{\dagger}e^{-i(k''.r'-\omega''t'-\l\p'_{k})}\hat{a}_{k''',\l}e^{i(k'''.r'-\omega'''t'-\l\p'''_{k})}\\&\quad+\hat{a}_{k,\l}^{\dagger}e^{-i(\textbf{k.r}-\omega t-\l\p_{k})}\hat{a}_{k',\l}e^{i(\bm{k}'.\textbf{r}-\omega't-\l\p'_{k})}\hat{a}_{k'',\l}e^{i(k''.r'-\omega''t'-\l\p'_{k})}\hat{a}_{k''',\l}^{\dagger}e^{-i(k'''.r'-\omega'''t'-\l\p'''_{k})}\Bigg]|n_{\xi,\l}\rangle.
\end{align*}

For simplicity, we define the constituent integrals in the expression above as:
\begin{align*}
    \beta_{1}&=\frac{1}{\sqrt{2\pi}}\int d^{3}k\xi_{l}(k)e^{i(\textbf{k.r}-\omega t-\l\phi_{k})},\\
    \beta_{2}&=\frac{1}{\sqrt{2\pi}}\int d^{3}k\rho_{k}\cos\phi_{k}\xi_{l}(k)e^{i(\textbf{k.r}-\omega t-\l\phi_{k})},\\
    \beta_{3}&=\frac{1}{\sqrt{2\pi}}\int d^{3}k\rho_{k}\sin\phi_{k}\xi_{l}(k)e^{i(\textbf{k.r}-\omega t-\l\phi_{k})},\\
    \beta_{4}&=\frac{1}{\sqrt{2\pi}}\int d^{3}k\Big(\rho_{k}\cos\phi_{k}\Big)^{2}\xi_{l}(k)e^{i(\textbf{k.r}-\omega t-\l\phi_{k})},\\
    \beta_{5}&=\frac{1}{\sqrt{2\pi}}\int d^{3}k\Big(\rho_{k}\sin\phi_{k}\Big)^{2}\xi_{l}(k)e^{i(\textbf{k.r}-\omega t-\l\phi_{k})}.
\end{align*}
Thus, the correlation of the longitudinal OAM density for a Fock-state in the limit ($r\rightarrow r'$) is given as:
\begin{equation}
    \langle n_{\xi,\l}|\hat{\mathcal{L}_{z}}^{2}(r,t)|n_{\xi,\l}\rangle=\frac{3n^{2}\hbar^{2}}{2(2\pi)^{3}}\left(x^{2}\beta_{1}^{*}\beta_{3}^{*}\beta_{3}\beta_{1}-xy\left(\beta_{1}^{*}\beta_{3}^{*}\beta_{2}\beta_{1}+\beta_{1}^{*}\beta_{2}^{*}\beta_{3}\beta_{1}\right)+y^{2}\beta_{1}^{*}\beta_{2}^{*}\beta_{2}\beta_{1}\right)
\end{equation}

Finally, we can write the fluctuations in the longitudinal OAM density along for a Fock-state pulse under paraxial approximation:
\begin{equation}
    \Delta\hat{\mathcal{L}}_z = \sqrt{\langle n_{\xi,\l}|\hat{\mathcal{L}_{z}}^{2}(r,t)|n_{\xi,\l}\rangle - \langle n_{\xi,\l}|\hat{\mathcal{L}_{z}}(r,t)|n_{\xi,\l}\rangle^{2}}
\end{equation}

Similarly, we can compute the OAM density fluctuations along the other directions for both Fock-state and coherent state pulses.

\section{Heisenberg uncertainty relations}

In this section, we show the validity of the Heisenberg uncertainty relations for the Fock and coherent photonic states. The Heisenberg relation is given by:

\begin{equation}
    \sqrt{\Delta \hat{L}_i^2 \Delta \hat{L}_j^2} \geq \frac{\hbar}{2}|\langle\hat{L}_k\rangle|,\quad i,j,k=k_x,k_y,k_z
\end{equation}

\begin{figure}[ht]
    \centering
    \includegraphics[width=0.7\textwidth]{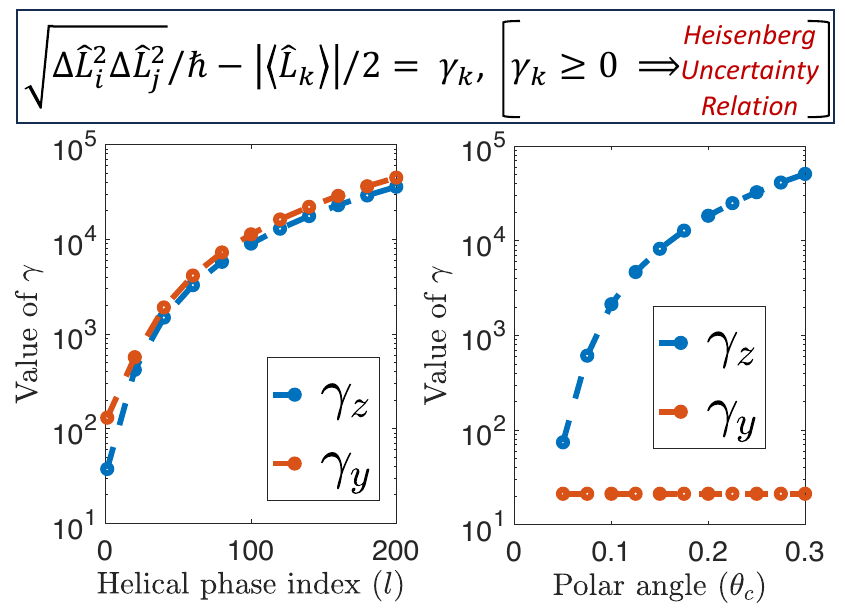}
    \caption{\textbf{Validation of Heisenberg uncertainty principle for photonic Fock state}. For coherent state, it is trivial from the plots since $\langle\hat{L}\rangle_{\text{Fock}} = \langle\hat{L}\rangle_{\text{coherent}}$, and $\Delta\hat{L}_{\text{coherent}}>\Delta\hat{L}_{\text{Fock}}$. We omitted the trivial case of $k=x$ since $|\langle\hat{L}_x\rangle|/2=0$. For the numerical calculations, we used $n=1$. Other parameters: $l=60$ for the $\theta_c$-dependent plot and $\theta_c=0.2\pi$ for the $l$-dependent plot.}
    \label{A4}
\end{figure}
\bigskip

\Urlmuskip=0mu plus 1mu\relax

\bibliographystyle{iopart-num}
\bibliography{biblio}

\end{document}